\documentclass[sigconf]{acmart}

\usepackage{booktabs} 

\usepackage{amsmath}
\usepackage[ruled]{algorithm2e}
\usepackage{subfigure}
\usepackage{graphicx}
\usepackage{diagbox}

\usepackage{bm}
\usepackage{xcolor}
\usepackage{soul}

\usepackage{multicol}

\usepackage{lipsum}

\newcommand\blfootnote[1]{%
  \begingroup
  \renewcommand\thefootnote{}\footnote{#1}%
  \addtocounter{footnote}{-1}%
  \endgroup
}

\copyrightyear{2017}
\acmYear{2017}
\setcopyright{acmcopyright} 
\acmConference{KDD'17}{}{August 13--17, 2017, Halifax, NS, Canada.} \acmPrice{15.00} 
\acmDOI{http://dx.doi.org/10.1145/3097983.3098122}
\acmISBN{978-1-4503-4887-4/17/08}

\begin{document}
\title[LSARS for Both Home-Town and Out-of-Town Users]{A Location-Sentiment-Aware Recommender System for Both Home-Town and Out-of-Town Users}

\author{Hao Wang*\blfootnote{*Corresponding authors; Yanmei Fu is a co-first author. 
}}
\affiliation{%
  \institution{Qihoo 360 Search Lab
}  
  \city{Beijing}
  \state{China}
}
\email{cashenry@126.com}

\author{Yanmei Fu*}
\affiliation{%
  \institution{Institute of Software, Chinese Academy of Sciences \\University of Chinese Academy of Sciences}
  \city{Beijing}
  \state{China}
}
\email{ustbfym3252@163.com}

\author{Qinyong Wang*}
\affiliation{%
  \institution{School of Information Technology and Electrical Engineering, The University of Queensland}
  \city{Brisbane}
  \state{Australia}
}
\email{wqinyongcas@163.com}

\author{Hongzhi Yin*}
\affiliation{%
  \institution{School of Information Technology and Electrical Engineering, The University of Queensland}
  \city{Brisbane}
  \state{Australia}
}
\email{h.yin1@uq.edu.au}

\author{Changying Du}
\affiliation{%
  \institution{Institute of Software, Chinese Academy of Sciences}
  \city{Beijing}
  \state{China}
}
\email{changying@iscas.ac.cn}

\author{Hui Xiong}
\affiliation{%
  \institution{Management Science and Information Systems Department, Rutgers University}
  \city{Newark}
  \state{USA}
}
\email{hxiong@rutgers.edu}

\begin{abstract}
Spatial item recommendation has become an important means to help people discover interesting locations, especially when people pay a visit to unfamiliar regions. Some current researches are focusing on modelling individual and collective geographical preferences for spatial item recommendation based on users' check-in records, but they fail to explore the phenomenon of user interest drift across geographical regions, i.e., users would show different interests when they travel to different regions. Besides, they ignore the influence of public comments for subsequent users' check-in behaviors. Specifically, it is intuitive that users would refuse to check in to a spatial item whose historical reviews seem negative overall, even though it might fit their interests. Therefore, it is necessary to recommend the \emph{right} item to the \emph{right} user at the \emph{right} location. In this paper, we propose a latent probabilistic generative model called LSARS to mimic the decision-making process of users' check-in activities both in home-town and out-of-town scenarios by adapting to user interest drift and crowd sentiments, which can learn location-aware and sentiment-aware individual interests from the contents of spatial items and user reviews. Due to the sparsity of user activities in out-of-town regions, LSARS is further designed to incorporate the public preferences learned from local users' check-in behaviors.
Finally, we deploy LSARS into two practical application scenes: spatial item recommendation and target user discovery. Extensive experiments on two large-scale location-based social networks (LBSNs) datasets show that LSARS achieves better performance than existing state-of-the-art methods.

\end{abstract}

%
%

\ccsdesc[300]{Information systems~Geographic information systems}
\ccsdesc[300]{Information systems~Data mining}
\ccsdesc[300]{Information systems~Social recommendation}


\keywords{Crowd sentiment; User interest drift; Check-in behavior; Recommendation}

\maketitle

\section{Introduction}
With the rapid development of wireless communication techniques,
the \emph{Location-Based Social Networks} (LBSNs) smart-phone
applications are constantly emerging, such as
Foursquare and Yelp, which help users find interesting places nearby. Besides,
users may also post their \emph{check-ins}, subjective reviews and images for places in these LBSNs platforms. Thus, a large number of users' behaviors in the physical world are recorded. How to deeply analyse users' behaviors for better personalized recommendations is a very valuable work.




Some existing methods apply collaborative filtering with matrix factorization for spatial item recommendation in LBSNs. In fact, there are millions of spatial items, but each user visits only a small number of them, resulting in data sparsity. To alleviate the problem, some researches fuse both geographical and social influences into matrix factorization \cite{cheng2011exploring,lian2014geomf}. However, previous methods are only appropriate for local users to recommendate spatial items, i.e., \textit{home-town recommendation}. In addition, for out-of-town users (e.g., travellers), most of their footprints are left in their home-town areas, which exacerbates the problem of data sparsity. \cite{hometown:h2} has reported that, for an average user,
the ratio of her home-town and out-of-town check-in records is 1:0.0047.




Leveraging the content information of spatial items is an effective
method to resolve the problems of data sparsity and cold start.
Some recent researches \cite{liu2013point,region:r,ye2010location,region:r2,lian2014geomf,ge2010energy,yin2015modeling} utilize the categories and tags of Points of Interests (POIs) to infer users' interests in out-of-town areas, and make POI recommendations accordingly. However, these methods fail to consider the phenomenon of \emph{user interest drift} \cite{yin2016adapting} across different geographical regions, i.e.,
users tend to have different interests when they travel out of their home town. For example, a user who rarely eats seafood
in her home town may frequently have a visit to the seafood restaurants
when she is travelling in a seaside city.
The reason behind the phenomenon is that users in out-of-town areas prefer to experiencing the local attractions which probably do not match their original interests.

\begin{figure}
\centering
\includegraphics[scale = 0.6]{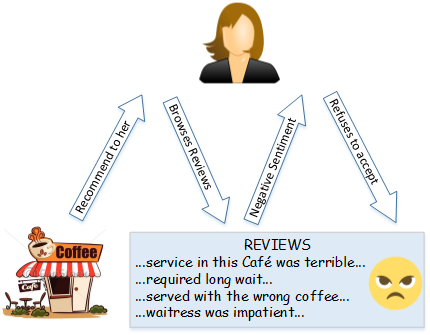}
\caption{An example of the decision-making process of a user visiting to a place}
\label{fig:process}
\end{figure}

Moreover, users usually express their subjective emotion for some visited places in their reviews.
Some literatures ~\cite{yang2013sentiment,gao2015content} have exploited the sentiment from historical reviews as and explicit feature to learn users' interests, but few researches take the crowd sentiment (i.e., the overall emotional tendency from all the reviews) for an item into account.
In fact, before users decide to visit a place, they usually refer to the relevant reviews for that place. That is, the decision-making process of users to visit a place is heavily influenced by previous users' attitudes. Thus, it is inappropriate to recommend a spatial item with negative sentiment to the users, even if the item itself match users' interests. Fig. \ref{fig:process} shows an example of the decision-making process of a user visiting to a place. The crowd sentiment for each item could be obtained by accumulating the sentiments from all the reviews.




%

Considering the influences of user interest drift and crowd sentiment, we propose a latent class probabilistic generative model called \emph{LSARS}, to simulate the decision-making process of users' check-in behaviours both in home-town and out-of-town areas. LSARS can adapt to users' dynamic interests in different areas and crowd sentiment towards each spatial item.

Inspired by existing methods on modelling user interest \cite{hu2013spatial,yin2016spatio}, LSARS employs latent topics to characterize users' interests so as to overcome the data sparsity. Specifically, we first divide the geographical space into some subregions where an individual's interest can be inferred over a set of topics. As regions and topics are interdependent, LSARS combines geographical clustering and topic modelling (i.e., Latent Dirichlet allocation (LDA) \cite{blei2003latent}) into a unified process. To further alleviate data sparsity of user behaviours, LSARS incorporates the crowd's preferences. To solve the problem of user interest drift, LSARS combines local specialties and user interests. Specifically, when user \textit{u} is in a region \textit{r} (i.e., home-town or out-of-town region),
LSARS would recommend the spatial items to users which are not only popular in \textit{r} but also satisfy \textit{u}'s interests. Besides, the attitudes from the crowd for spatial items strongly affect users' behaviors, thus LSARS employs a sentiment-LDA model. Moreover, LSARS utilizes the Tobler's First Law of geography \cite{miller2004tobler} to model users' interests, i.e., if two regions are geographically proximate, then the users' interests in these two regions should be similar.

%
%
%

To the best of our knowledge, this is the first work to model users' check-in activities in home-town and out-of-town regions by adapting to both user interest drift and crowd sentiment for spatial items. Two corresponding components are built in LSARS: \emph{Popularity-Aware User Mobility}(PUM) and \emph{Sentiment-Aware User Interest}(SUI).

\begin{table}[b]
\small
\newcommand{\tabincell}[2]{\begin{tabular}{@{}#1@{}}#2\end{tabular}}
\centering

\setlength{\belowcaptionskip}{8pt}
\caption{Notations of the Input Data}
\label{tb:Notations of the Input Data}
\begin{tabular}{c|c} \hline
SYMBOL&DESCRIPTION\\ \hline
$N,V,M,W,C$&\tabincell{c}{the number of users, spatial items, \\locations, content words, review words}\\ \hline
$D_u$ & the profile of user $u$\\ \hline
$v_{u,i}$ & the spatial item of $i^{th}$ record in $D_u$ \\ \hline
$l_{u,i}$ & the location of spatial item $v_{u,i}$ \\ \hline
$l_u$ & the home location of user $u$\\\hline
$W_{u,i}$ & set of words describing spatial item $v_{u,i}$ \\ \hline
$C_{u,i}$ & \tabincell{c}{set of words of $u$'s review about $v_{u,i}$} \\ \hline

$\alpha,\gamma,\beta,\eta,\delta,\tau$ & \tabincell{c}{Dirichlet priors to multinomial distribution\\

$\theta_u,\vartheta_u,\phi_{zs},\varphi_z,\omega_z,\varphi_r$} \\ \hline
\end{tabular}
\end{table}

To demonstrate the effectiveness of LSARS, we apply it in two application scenarios: \emph{spatial item recommendation} and \emph{target user discovery}. In each scenario, we consider both home-town and out-of-town cases. The main
contributions can be summarized as follows:
\begin{itemize}

\item \textbf{Sentiment Influence:} we model sentiment influence from public reviews into user interests. Before users decide to check in to a place, they not only check the content information of that place but also pay more attention to user reviews.

\item \textbf{Algorithm and Model:} we propose a latent class probabilistic generative model called LSARS which can accurately capture users' check-in behaviors by considering user preferences, geographical influences, content effects and sentiment influences in a unified way.

\item \textbf{Empirical Evaluations:} we conduct extensive experiments to evaluate the performance of the proposed LSARS model in two applications of spatial item recommendation and target user discovery, by using two large-scale datasets from LBSN platforms, and the results demonstrate our approach outperforms state-of-the-art baselines.

\end{itemize}


The rest of this paper is organized as follows. Section 2 introduces the preliminaries and problem formulations. Section 3 describes the proposed LSARS model, and then presents the inference algorithm. Section 4 illustrates two applications of  LSARS separately in spatial item recommendation and target user discovery. The experimental results are reported in Section 5. Section 6 reviews the related work and Section 7 concludes this paper.

\section{Preliminaries and Problem Formulation}
In this section, we first define relevant data structures and notations used in this paper, and then formulate the problems. For ease of presentation, Table \ref{tb:Notations of the Input Data} lists the notations of the input data.

\emph{Definition 1.} \textbf{(Spatial Item)} A spatial item is an item associated with a geographical location (e.g., a restaurant or a cinema). In our model, a spatial item has three attributes: identifier, location and content. We use $v$ to represent a spatial item identifier and $l_v$ to denote its corresponding geographical attribute with longitude and latitude coordinates. Besides, a spatial item includes textual semantic information, such as categories and tags. We use the notation $W_v$ to denote the set of words describing the spatial item $v$.

\emph{Definition 2.} \textbf{(Check-in Activity)} A user's check-in activity is represented by a five tuple ($u,v,l_v,W_v,C_v$) which indicates that the user $u$ visits the spatial item $v$ with the content description $W_v$ and user reviews $C_v$ at the location $l_v$.

\emph{Definition 3.} \textbf{(User Home Location)} Given a user $u$, we denote $l_u$ as the user's home location where the user lives in. However, it is hard to directly obtain a user's home location. Thus, we use the method developed by \cite{hometown:h2},
such that we regard the spatial item where a user frequently checks in as her home location.

\emph{Definition 4.} \textbf{(User Profile)} For each user $u$, we create a user profile $D_u$, which is a set of user check-in activities associated with $u$. In fact, the dataset $D$ is a collection of user profiles, $D = \{ D_u: u\in{U}\}$.




Given the dataset $D$, the first target is to provide spatial item recommendation for both home-town and out-of-town users. Furthermore, the owners of spatial items would also like to discover potential users. Thus, the second target is to recommend both home-town and out-of-town users to the owner of each spatial item. Therefore, we formulate our problems with consideration of two different scenarios in a unified way as follows.

PROBLEM 1. \textbf{(\emph{Spatial Item Recommendation)}} Given users' check-in dataset $D$ and a user $u$ with his/her current location $l$, our goal is to recommend a list of spatial items that $u$ may be interested in (that is, the query is $q = (u,l)$). Given a distance threshold $d$, if the distance between current location and home location (that is, $|l-l_u|$) of a user is larger than $d$, it is in an \textbf{out-of-town recommendation}. Otherwise, it is a \textbf{home-town recommendation}.

PROBLEM 2. \textbf{(\emph{Target User Discovery})} Given a spatial item $v$ at $l_v$, our goal is to discover a list of target users who may favor the spatial item $v$. Given a distance threshold $d$, it becomes an \textbf{out-of-town user discovery} if the distance between the position of the spatial item and the home location of the target user (that is,
$|l_v-l_u|$) is greater than $d$. Otherwise, it is a \textbf{home-town user discovery}.

In line with \cite{distance:d,distance:d2}, we set $d = 100km$ in this paper, which takes at least one hour to drive, because the distance of around $100km$ is the typical range of human daily physicial activity, 

\begin{table}

\newcommand{\tabincell}[2]{\begin{tabular}{@{}#1@{}}#2\end{tabular}}
\centering

\setlength{\belowcaptionskip}{8pt}
\small
\caption{Notations of model parameters}
\label{tb:Notations of Model Parameters}

\begin{tabular}{c|c} \hline
SYMBOL&DESCRIPTION\\ \hline
$S,R,K$&\tabincell{c}{the number of sentiments, regions, topics}\\ \hline
$s_{u,i}$ & \tabincell{c}{the sentiment of user review about $v_{u,i}$} \\ \hline
$s_+, s_-$ & \tabincell{c}{the positive sentiment, negative sentiment } \\ \hline
$\mu_r$ & \tabincell{c}{the mean location of region $r$} \\ \hline
$\Sigma_r$ & \tabincell{c}{the location covariance of region $r$} \\ \hline
$\vartheta_u$ & \tabincell{c}{$u$'s check-in activity range, distribution over regions} \\ \hline
$\theta_u$ & \tabincell{c}{the interests of user $u$, distribution over topics} \\ \hline
$\varphi_r$ & \tabincell{c}{$r$'s region-level popularity distribution over spatial items} \\ \hline
$\psi_z$ & \tabincell{c}{topic $z$'s distribution over words for spatial items} \\ \hline
$\omega_z$ & \tabincell{c}{topic $z$'s distribution over sentimental words} \\ \hline
$\phi_{zs}$ & \tabincell{c}{distribution over words of user's reviews} \\ \hline
\end{tabular}
\end{table}

\section{The LSARS Model}
\subsection{Model Description}

To model the decision-making process of a user visiting a spatial item both in home-town and out-of-town regions, we propose a joint probabilistic generative model called LSARS, which assumes that the check-in behavior of a user is influenced by the following factors: geographical influence, item content effect and user review effect. Table \ref{tb:Notations of the Input Data} shows the graphical representation of LSARS and Table \ref{tb:Notations of Model Parameters} lists relevant notations of the model. In LSARS, users' check-in records are modeled as observed random variables while the topic, region and sentiment are considered as latent random variables, which are denoted as $z$, $r$ and $s$, respectively. Specifically, LSARS includes two components: Sentiment-Aware User Interest and Popularity-Aware User Mobility.\\

\begin{figure}
\centering
\includegraphics[scale = 0.95]{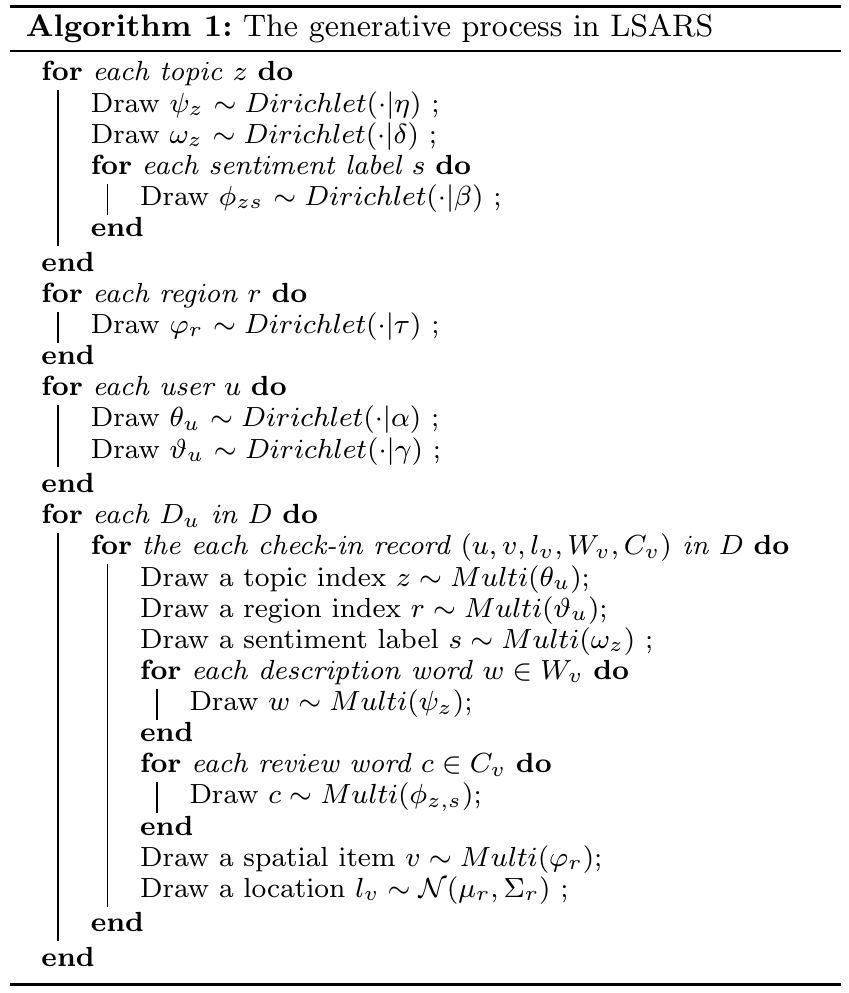}
\end{figure}

\textbf{Sentiment-Aware User Interest Modelling.} Intuitively, a user finally chooses a spatial item based on the following assumptions: 1) the content of the spatial item matching his/her personal interests and 2) the item receiving the most positive reviews.

Inspired by the early work on user interest modelling \cite{hu2013spatial,topic:t2}, LSARS adopts latent semantic topics to characterize a user's interests. Specifically, we infer a user's interest distribution over a set of topics in terms of the contents (e.g., tags and categories) of his/her check-in spatial items, denoted as $\theta_u$.



We also model the user reviews to infer their preferences. Technically, each topic $z$ in our model is not only associated with a multinomial distribution over words $\psi_z$ of item contents, but also related with a multinomial distribution over words $\phi_{zs}$ of user reviews.
In order to obtain the crowd sentiment for an item, LSARS accumulates the sentiments from all the reviews for this item
as the model parameters $\phi$ and $\varphi$ are shared by all the users.  Following the work in \cite{sentiment:s}, we integrate user reviews with latent sentiments to the topic discovery process and employ a multinomial distribution over sentiment words $\omega_z$. The review word  $c_v$ is characterized by $P(c_v|\bm{z},\bm{s},\beta)$ as follows:


\begin{equation}
\small
\label{eq:pr_cv}
 P(c_v|\bm{z},\bm{s},\beta) = (\frac{\Gamma(\sum_c\beta_c)}{\prod_c\Gamma(\beta_c)})^{K*S}\prod_z\prod_s\frac{\prod_c\Gamma(n_{z,s,c}+\beta_c)}{\Gamma(\sum_c(n_{z,s,c} +\beta_c))}
\end{equation}

where $\Gamma(.)$ is the gamma function. To avoid overfitting, we place a Dirichlet prior, parameterized by $\beta$, over the multinomial distribution $\phi_{zs}$. Similarly, priors over $\theta_u,\vartheta_u,\omega_z,\varphi_r,\psi_z$ are imposed with parameters $\alpha,\gamma,\delta,\tau,\eta$.


Note that in a typical topic model such as \cite{blei2003latent}, a document contains multiple topics and each word has a latent topic label, which is suitable for dealing with long text. However, for short text $W_v$ or $C_v$, a document is more likely to contain only a single topic. Therefore, in LSARS, all the words in $W_v$ or $C_v$ are assigned with a single topic $z$.

\textbf {Location-Aware User Mobility Modelling.} Users are more likely to visit a number of spatial items within some geographical regions. Thus we divide the geographical space into $R$ regions and employ a multinomial distribution $\vartheta_u$ over regions to model $u$'s spatial patterns. In line with \cite{region:r} and \cite{region:r2}, we apply a Gaussian distribution for each region $r$, and the location for a spatial item $v$ is characterized by $l_v \sim \mathcal{N}(\mu_r,\Sigma_r)$ as follows:

\begin{equation}
\small
\label{eq:l_v}
P(l_v|\bm{\mu_r},\bm{\Sigma_r}) = \frac{1}{2\pi\sqrt{|\bm{\Sigma_r}|}}\exp(\frac{-(l_v - \bm{\mu_r})^K\bm{\Sigma_r}^{-1}(l_v - \bm{\mu_r})}{2})
\end{equation}
where \bm{$\mu_r$} and \bm{$\Sigma_r$} denote the mean vector and covariance matrix, respectively.

Popularity also has a great effect on users' check-in activities, especially when users are in out-of-town areas. Specifically, users' check-in decisions are strongly affected by the popularity of spatial items. We use a multinomial distribution $\varphi_r$ to model the popularity of spatial items in a region level.
This is a key model design that enables both home-town and out-of-town recommendations.


Finally, we obtain the joint distribution of the observed and hidden variables in Equation (\ref{eq:joint-geolda}).
\begin{equation}
\scriptsize
\label{eq:joint-geolda}
\begin{split}
&P(\boldmath{v},\boldmath{w},\boldmath{c_v},\boldmath{l_v},\boldmath{s},\boldmath{z},\boldmath{r}|\alpha,\beta,\gamma,\tau,\eta,\delta,\boldmath{\Sigma},\boldmath{\mu})\\
&=P(\boldmath{z}|\alpha)P(\boldmath{r}|\gamma)P(\boldmath{w}|\eta,\boldmath{z})P(\boldmath{s}|\delta,\boldmath{z})P(\boldmath{c_v}|\boldmath{z},\boldmath{s},\beta)P(\boldmath{v}|\boldmath{r},\tau)P(\boldmath{l_v}|\boldmath{r},\boldmath{\mu},\boldmath{\Sigma})\\
&=\int\ldots\int P(\boldmath{z}|\theta)P(\theta|\alpha) P(\boldmath{r}|\vartheta)P(\vartheta|\gamma) P(\boldmath{w}|\boldmath{z},\psi)P(\psi|\eta) P(\boldmath{s}|\boldmath{z},\omega)\\
&P(\omega|\delta)P(\boldmath{c_v}|\boldmath{z},\boldmath{s},\phi)P(\phi|\beta) P(\boldmath{v}|\boldmath{r},\varphi)P(\varphi|\eta) P(\boldmath{l_v}|\boldmath{r},\boldmath{\mu},\boldmath{\Sigma}) d\theta d\vartheta  d\omega d\phi  d\psi d\varphi
\end{split}
\end{equation}

\begin{figure}
\centering
\includegraphics[scale = 0.60]{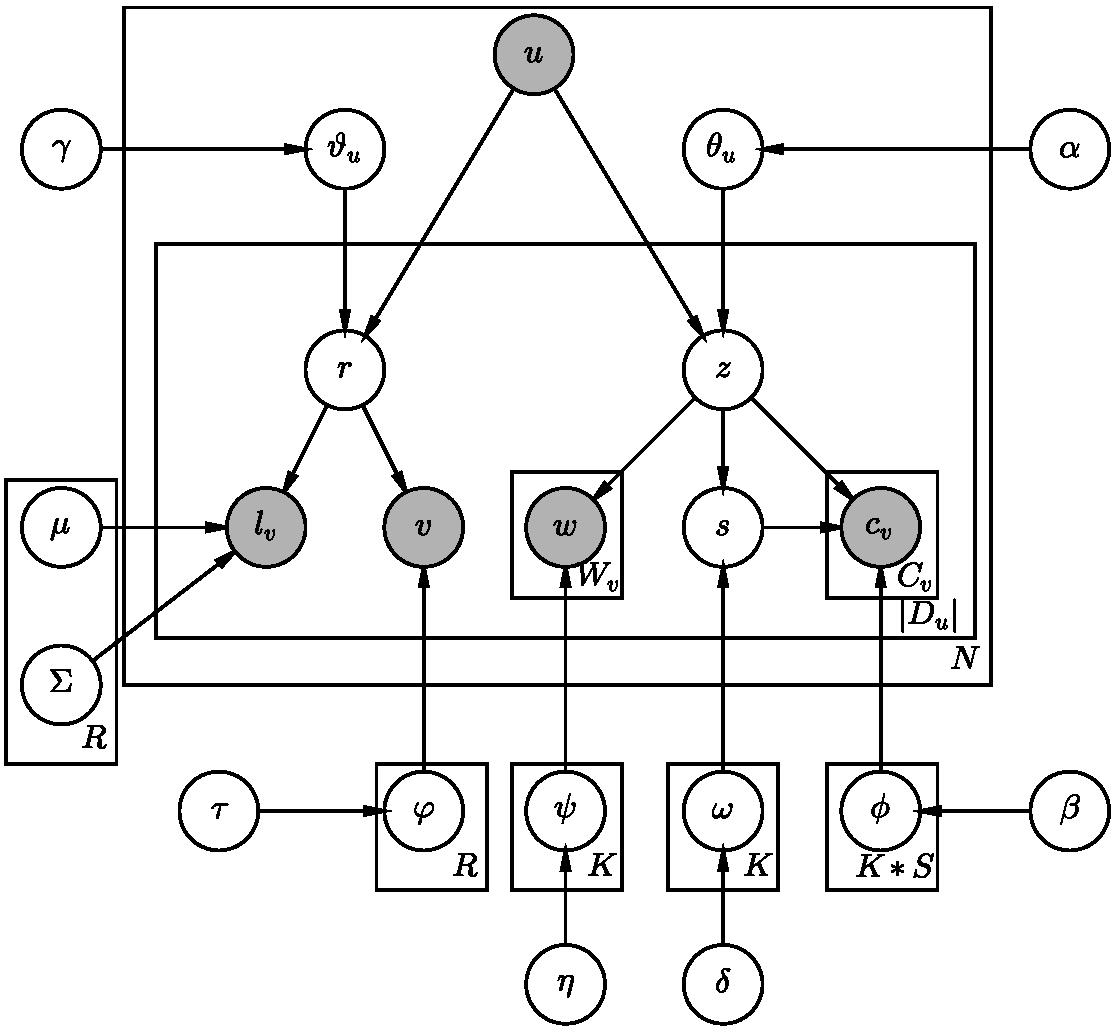}
\caption{The Graphical Representation of LSARS}
\label{fig:model}
\end{figure}

The probabilistic generative process of LSARS model is listed in Algorithm 1 and the graphical representation of LSARS is shown in Figure~\ref{fig:model}.

\subsection{Model Inference}

Given the hyper-parameters $\alpha,\beta,\gamma,\eta,\delta,\tau$ and the observations \bm{$v,l_v,W_v$} and \bm{$C_v$}, our goal is to infer the latent variables $\theta,\vartheta,\phi,\psi,\varphi$ and $\omega$, as well as the parameters $\mu$ and $\Sigma$. For $\theta,\vartheta,\phi,\psi,\varphi$ and $\omega$, we use the posterior distribution over latent topic $z$, latent sentiment $s$ and latent region $r$ for each user, instead of explicitly representing them as parameters to be estimated, 
As it is intractable to exactly computing this posterior, we adopt the Markov Chain Monte Carlo (MCMC) method to draw samples, as in \cite{gibbs:g,gibbs:g2}.

In MCMC, a Markov chain is constructed to converge to the target distribution, and samples are then taken
from that Markov chain. Each state of the chain is an assignment of values to the variables being sampled, and different transition rules between states need different samplers.
Thus gibbs sampling is used. First, the conditional probability of $z$ and $s$: $P(z,s|\bm{z_{\neg{u},v}},\bm{s_{\neg{u},v}},\bm{v},\bm{r},\bm{l_v},\bm{W_v},\bm{C_v},u)$ should be estimated.
The joint probability distribution of the latent and observed variables is shown in Equation (\ref{eq:joint-geolda}), and using the Bayes chain rule, the conditional probability is obtained as follows:


\begin{equation}
\small
\label{eq:pr_zs}
\begin{split}
&P(z,s|\bm{z_{\neg{u}},\bm{v}},\bm{s_{\neg{u}},},\bm{v},\bm{r},\bm{l_v},\bm{W_v},\bm{C_v},u) \propto \frac{n_{u,z}^{\neg{u},v}+\alpha_z}{\sum_{z'}(n_{u,z'}^{\neg{u},v}+\alpha_{z'})} \\
& \times \prod_{w\in{W_v}}\frac{n_{z,w}^{\neg{u},v}+\eta_w}{\sum_{w'}(n_{z,w'}^{\neg{u},v}+\eta_{w'})}
\frac{n_{z,s}^{\neg{u},v}+\delta_s}{\sum_{s'}(n_{z,s'}^{\neg{u},v}+\delta_{s'})} \\
& \times \prod_{c\in{C_v}}\frac{n_{z,s,c}^{\neg{u},v}+\beta_c}{\sum_{c'}(n_{z,s,c'}^{\neg{u},v}+\beta_{c'})}
\end{split}
\end{equation}
where $n_{u,z}$ is the number of times that latent topic $z$ has been sampled from user $u$; $n_{z,w}$ is the number of times that word $w$ is generated from topic $z$; $n_{z,s}$ is the number of times that sentiment $s$ is generated from topic $z$; $n_{z,s,c}$ is the number of times that word $c_v$ is generated from topic $z$ with sentiment label $s$. The number $n^{\neg{u,v}}$ with superscript $\neg{u,v}$ denotes a quantity excluding the current instance.

Then, we sample region $r$ according to the following posterior probability:
\begin{equation}
\small
\label{eq:pr_r}
\begin{split}
&P(r|\bm{r_{\neg{u}},\bm{v}},\bm{z},\bm{s},\bm{l_v},\bm{W_v},\bm{C_v},u)\\
&\propto \frac{n_{u,r}^{\neg{u},v}+\gamma_r}{\sum_{r'}(n_{u,r'}^{\neg{u},v}+\gamma_{r'})}
\frac{n_{r,v}^{\neg{u},v}+\tau_v}{\sum_{v'}(n_{r,v'}^{\neg{u},v}+\tau_{v'})}P(l_v|\mu_r,\Sigma_r)
\end{split}
\end{equation}
where $n_{u,r}$ is the number of times that region $r$ has been sampled from user $u$, and $n_{r,v}$ is the number of times that spatial item  $v$ is generated by region $r$.

After each iteration, the parameters $\mu_r$ and $\Sigma_r$ are updated as in Equation (\ref{eq:mu_r}) and (\ref{eq:sigma_r}).

\begin{equation}
\small
\label{eq:mu_r}
\bm{ \mu_r} = E(r) = \frac{1}{|S_r|}\sum_{v \in{S_r}}l_v
\end{equation}
\begin{equation}
\small
\label{eq:sigma_r}
\bm{\Sigma_r} = D(r) = \frac{1}{|S_r|-1}\sum_{v \in{S_r}}(l_v-\bm{\mu_r})(l_v-\bm{\mu_r})^K
\end{equation}
	where $S_r$ denotes the set of spatial items assigned with latent region $r$.
	
\textbf{Inference Framework.} For simplicity, we fix the hyper-parameters as $\alpha = \frac{50}{K}$, $\gamma = \frac{50}{R}$ , $\beta=\eta=\tau=\delta=0.01$ in our implementation. We initialize the latent geographical region $r$ by a k-means clustering algorithm, and then randomly initialize topic $z$ and sentiment $s$ assignments for check-in records. And then, in each iteration, we use Equation (\ref{eq:pr_zs}) and (\ref{eq:pr_r}) to update the region, topic and sentiment assignments. After each iteration, Equation (\ref{eq:mu_r}) and (\ref{eq:sigma_r}) are used to update the Gaussian distribution parameters. The iteration is repeated for 1600 times. Finally, the posterior samples can be used to estimate parameters by examining the counts of $z$ and $r$ assignments for check-in records. The parameters $\vartheta$, $\theta$, $\varphi$, $\phi$, $\psi$ and $\omega$ are then estimated as follows:

\begin{equation}
\small
  \label{eq:theta}
     \hat{{\theta}}_{u,z} = \frac{n_{u,z}+\alpha_z}{\sum_{z'}(n_{u,z'}+\alpha_{z'})};
     \hat{{\vartheta}}_{u,r} = \frac{n_{u,r}+\gamma_r}{\sum_{r'}(n_{u,r'}+\gamma_{r'})}
\end{equation}
\begin{equation}
\small
\label{eq:omega}
     \hat{{\omega}}_{z,s} = \frac{n_{z,s}+\delta_s}{\sum_{s'}(n_{z,s'}+\delta_{s'})};
     \hat{{\psi}}_{z,w} = \frac{n_{z,w}+\eta_w}{\sum_{w'}(n_{z,w'}+\eta_{w'})}
\end{equation}
\begin{equation}
\small
	\label{eq:varphi}
     \hat{{\varphi}}_{r,v} = \frac{n_{r,v}+\tau_{v}}{\sum_{v'}(n_{r,v'}+\tau_{v'})};
     \hat{{\phi}}_{z,s,c} = \frac{n_{z,s,c}+\beta_{c}}{\sum_{c'}(n_{z,s,c'}+\beta_{c'})}
\end{equation}


\subsection{Time Complexity}
 We analyze the time complexity of the inference framework. Suppose that the whole process runs $I$ iterations. In each iteration, all the users' check-in records are scanned. For each check-in record, it requires $\mathcal{O}(KS)$ operations to compute the posterior distribution for sampling latent topic and latent sentiment, and requires $\mathcal{O}(R)$ operations to compute the posterior distribution for sampling latent region. Thus, the whole time complexity is $\mathcal{O}(I(KS+R)\sum_u|D_u|)$.

\section{Applications Using LSARS}
In this section, we deploy LSARS to two applications: spatial item recommendation and target user discovery.

\subsection{Spatial Item Recommendation}
Given a querying user $u_q$ and location $l_q$, i.e., $q = (u_q,l_q)$, the task is to compute a probability of a user $u_q$ checking in each spatial item $v$, and then return top-$k$ items with higher probability to the user $u_q$. Specifically, the probability of a user $u_q$ visiting spatial items $v$ is computed by Equation (\ref{eq:pr_item_v}), where we denote the positive sentiment as $s_{q+}$.


\begin{equation}
\small
\label{eq:pr_item_v}
P(v,l_v,W_v|u_q,l_q,s_{q+})
\propto P(v,l_v,W_v,s_{q+}|u_q,l_q)
\end{equation}
where $P(v,l_v,W_v,s_{q+}|u_q,l_q)$ is calculated as follows:

\begin{equation}
\small
\label{eq:pr_item_v_p}
P(v,l_v,W_v,s_{q+}|u_q,l_q)= \sum_rP(r|l_q)P(v,l_v,W_v,s_{q+}|u_q,r)
\end{equation}
where $P(r|l_q)$ denotes the probability of user $u$ visiting region $r$ given his/her current location $l_q$, and it is computed by Equation (\ref{eq:pr_latent_r}) according to Bayes rule. The prior probability of latent region $r$ can be estimated by Equation (\ref{eq:pr_prior_r}).
\begin{equation}
\small
\label{eq:pr_latent_r}
\begin{split}
P(r|l_q) = \frac{P(r)P(l_q|r)}{\sum_{r'}P(r')P(l_q|r')}
\propto P(r)P(l_q|r)
\end{split}
\end{equation}

\begin{equation}
\small
\label{eq:pr_prior_r}
\begin{split}
P(r) = \sum_uP(r|u)P(u) = \sum_u\frac{N_u + \kappa}{\sum_{u'}(N_{u'} + \kappa)}\vartheta_{u',r}
\end{split}
\end{equation}
where $N_u$ denotes the number of check-ins generated by user $u$. To avoid overfitting, we introduce the Dirichlet prior parameter $\kappa$ as the pseudo-count. We adopt the geometric mean as the probability of topic $z$ generating word set $W_v$.

\begin{equation}
\small
\label{pr_choose_v}
\begin{split}
&P(v,l_v,W_v,s_{q+}|u_q,r) = P(l_v|r)P(v|r)\sum_zP(z|u_q)P(s_{q+}|z)\\
&\times\prod_{w \in W_v}P(w|z)^{\frac{1}{|W_v|}}
\end{split}
\end{equation}

Based on Equations (\ref{eq:pr_item_v}) - (\ref{pr_choose_v}), we bring up Equation (\ref{eq:top_k}) to infer the score of each spatial item and provide top-$k$ items for users.

\begin{equation}
\small
\label{eq:top_k}
\begin{split}
&P(v,l_v,W_v,s_{q+},l_q) =\sum_r[P(r)P(l_v|\mu_r,\Sigma_r)P(l_q|\mu_r,\Sigma_r)\varphi_{r,v}] \\
&\times \sum_z[\theta_{u_q,z}\omega_{z,s_+} \prod_{{w \in W_v}}\psi_{z,w}^\frac{1}{|W_v|}]
\end{split}
\end{equation}

\subsection{Target User Discovery}
We apply our model to discover potential users for a spatial item $v$. Given a spatial item $v$ and its description $W_v$, the task is to find top-$k$ users who probably have a positive sentiment (denoted as $s_+$) for the item $v$. Specifically, the probability of a user $u$ visiting the spatial item $v$ is computed by considering both geographical influence and sentiment effect of the item:


\begin{equation}
\small
\label{eq:pr_user_p}
P(u,s_+|v,W_v) = \frac{P(u,s_+,v,W_v)}{\sum_u\sum_sP(u,v,s,W_v)}
\end{equation}
\begin{equation}
\small
\label{eq:pr_joint}
P(u,s,v,W_v) = P(u)\sum_zP(z|u)P(s|z)P(W_v|z)\sum_rP(v|r)
\end{equation}
where the prior $P(u)$ of users is calculated by Equation (\ref{eq:prior_user}):
\begin{equation}
\small
\label{eq:prior_user}
P(u) = \frac{N_u + \kappa}{\sum_{u'}(N_{u'} + \kappa)}
\end{equation}

\section{Experiments}

In this section, we first describe the settings of experiments and then demonstrate the experimental results.

\subsection{Datasets}

Our experiments are conducted on two real-world datasets: Yelp and Foursquare. The details of these two datasets are shown in Table \ref{tb:basic statistics}.

\begin{table}
\setlength{\belowcaptionskip}{8pt}
\caption{Basic statistics of Yelp and Foursquare datasets}
\label{tb:basic statistics}
\newcommand{\tabincell}[2]{\begin{tabular}{@{}#1@{}}#2\end{tabular}}
\centering

\begin{tabular}{c|c|c} \hline
&Yelp&Foursquare\\ \hline
\# of the users& 4,157& 17,417 \\ \hline
\# of the items& 27,696& 78,193 \\ \hline
\# of the check-ins& 157,657& 268,089 \\ \hline
time span & Apr 2009-Jan 2015& Jan 2009-Dec 2011  \\ \hline
\end{tabular}

\end{table}

\textbf{Yelp.} The Yelp's Challenge Dataset\footnote{$https://www.yelp.com/dataset\_challenge$} contains 4157 users who have at least 20 reviews and 27696 spatial items from four cities. Each check-in record is stored as \emph{user-ID}, \emph{Item-ID}, \emph{Item-location}, \emph{Item-content}, \emph{Item-review} and \emph{check-in date}. Note that this dataset does not contain the exact check-in time, and only provides the coarse check-in date (e.g., ``2010-01-01").


\textbf{Foursquare.} The Foursquare dataset\footnote{$http://www.public.asu.edu/hgao16/dataset.html$} contains the check-in records of 17417 users who post at least 6 reviews respectively in two cities (i.e., New York and Los Angels). Each check-in record is stored as \emph{user-ID}, \emph{Item-ID}, \emph{Item-location}, \emph{Item-content}, \emph{Item-review} and \emph{check-in time}.

\subsection{Comparative Approaches}

We compare LSARS with the following five state-of-the-art methods with well-tuned parameters.

\textbf{CKNN:} CKNN \cite{CKNN} projects a user's activities into the category space and models user preferences using a weighed category hierarchy. When receiving a query, CKNN retrieves all the users and items located in the querying area, and formulates a user-item matrix. Then it applies a user-based collaborative filtering (CF) method to predict the query user's rating for an unvisited item. The similarity between two users is computed by their weights in the category hierarchy.

\textbf{UPS-CF.} UPS-CF \cite{distance:d} is a collaborative recommendation framework for out-of-town users by considering user preference, social influence and geographical proximity. Specifically, UPS-CF recommends the spatial items to a target user based on the check-in records of both his/her friends and similar users.

\textbf{CAPRF.} CAPRF \cite{gao2015content} is a unified POI recommendation framework that integrates
sentiment indications, user interests, and POI properties, in which a novel sentiment-enhanced weighting scheme is proposed to incorporate personal sentiment information.

\textbf{GCF.} GCF \cite{ye2011exploiting} is a collaborative filtering model incorporating geographical influence.

\textbf{LCA-LDA.} LCA-LDA \cite{topic:t2} is a location-content-aware recommendation model which supports spatial item recommendation for out-of-town users. LCA-LDA considers both personal interests and local preferences in each city by exploiting the co-visiting pattern and the content of the spatial items. Compared with our LSARS, LCA-LDA fails to model the geographical influence
and crowd  sentiment.

\textbf{JIM.} JIM \cite{region:r} is a joint probabilistic generative model to mimic users' check-in behaviours by integrating the factors of temporal effect, content effect, geographical influence and word-of-mouth effect, especially for out-of-town users. Compared with our LSARS, JIM also fails to take crowd sentiment into account but incorporates the temporal effect instead.

\subsection{Evaluation Methods and Metrics}
In this section, we introduce the evaluation methods and metrics for the two applications of LSARS model.

\subsubsection{Evaluation for spatial item recommendation}

As LSARS is designed for both home-town and out-of-town recommendation, we evaluate the recommendation effectiveness on spatial item recommendation under these two scenarios respectively. To determine whether a user's activity occurs in the home-town region or not,
we measure the location distance between the home location $l_u$ of the user and the visiting item location $l_v$. If the distance (i.e., $|l_u -l_v|$) is greater than $d$,
then the user is considered to be in out-of-town region. Following previous researches \cite{yin2016spatio,method:m2}, we select $d$ as $100km$ in this paper.

To make an overall evaluation of the recommendation effectiveness, we follow the methodological framework proposed in \cite{region:r,hu2013spatial,method:m2} to compute $Accuracy@k$.
We split the user activity dataset $D$ into the training set $D_{train}$ and the test set $D_{test}$ (70\% for training and 30\% for testing).
Specifically, for each user activity record $(u,v,l_v,W_v,C_v)$ in $D_{test}$,
1) we compute the ranking score for spatial item
$v$ and all other unvisited items;
2) we obtain a ranked list by ordering all the spatial items; Let $p$ denote the position of spatial item $v$ in the list. The best result corresponds to the case where $v$ precedes all the unvisited items;
3) we make a top-$k$ recommendation list by picking up $k$ ranked spatial items from the list.
If $p \leq k$, we have a hit (i.e., the ground-truth visited item $v$ is recommended to the user); otherwise, we have a miss.

The computation of $Accuracy@k$ proceeds as follows. We define the value of $hit@k$ for a single test case as either 1, if the ground-truth spatial item $v$ appears in the top-$k$ results, or 0 if otherwise. The overall $Accuracy@k$ is defined by averaging over all the test cases:
\begin{equation}
\small
Accuracy@k = \frac{\#hit@k}{|D_{test}|}
\end{equation}
where $\#hit@k$ denotes the number of hits in the test set, and $|D_{test}|$ is the number of all the test cases.

\subsubsection{Evaluation for target user discovery}

Similarly, we carry out our model to discover potential users under the two scenarios respectively. We also adopt the same training and testing datasets.
We take three steps as below:
1) for each spatial item $v$, we compute $P(u,s_+|v,C_v)$ as the probability of user $u$'s checking in to the spatial item $v$;
2) we obtain a ranked list by ordering the value of $P(u,s_+|v,C_v)$ of all the users in descending order,
and pick the top $k$ users from the list;
3) we define the measurement metric as $Precision@k$:
\begin{equation}
\small
Precision@k = \frac{\#relevances}{k}
\end{equation}
where $\#relevances$ is the number of target users in the $top$-$k$ recommended users.

\subsection{Performance Analysis}
In this section, we show the experimental results in the tasks of spatial item recommendation and target user discovery with well-tuned parameters.
\subsubsection{Effectiveness of spatial item recommendation }
\begin{figure}
\small
\centering
\subfigure[Home-town spatial item recommendation]{
\label{fig:subfig:a} 
\includegraphics[width=0.45\columnwidth]{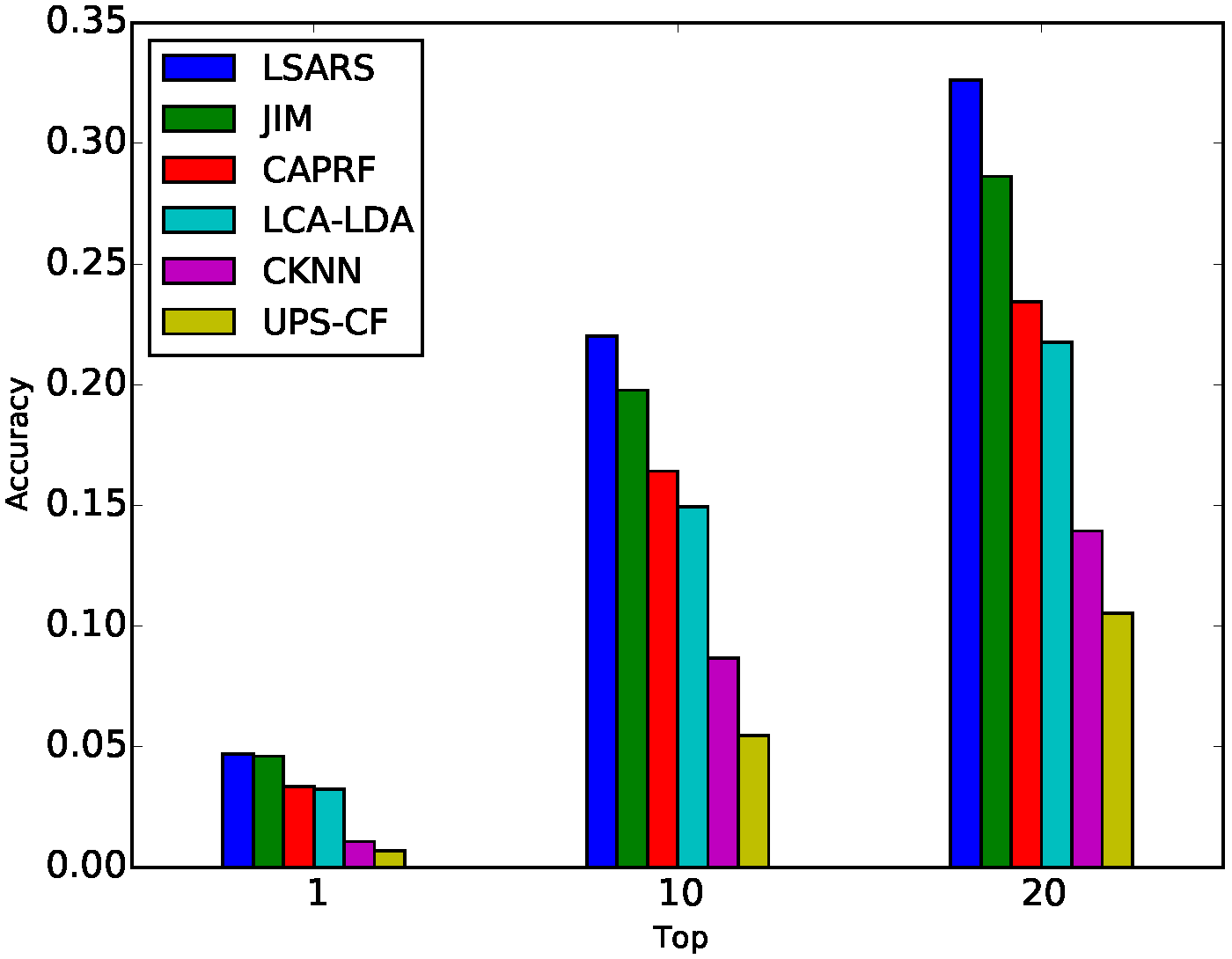}}
\hspace{0.1in}
\centering
\subfigure[Out-of-town spatial item recommendation]{
\includegraphics[width=0.45\columnwidth]{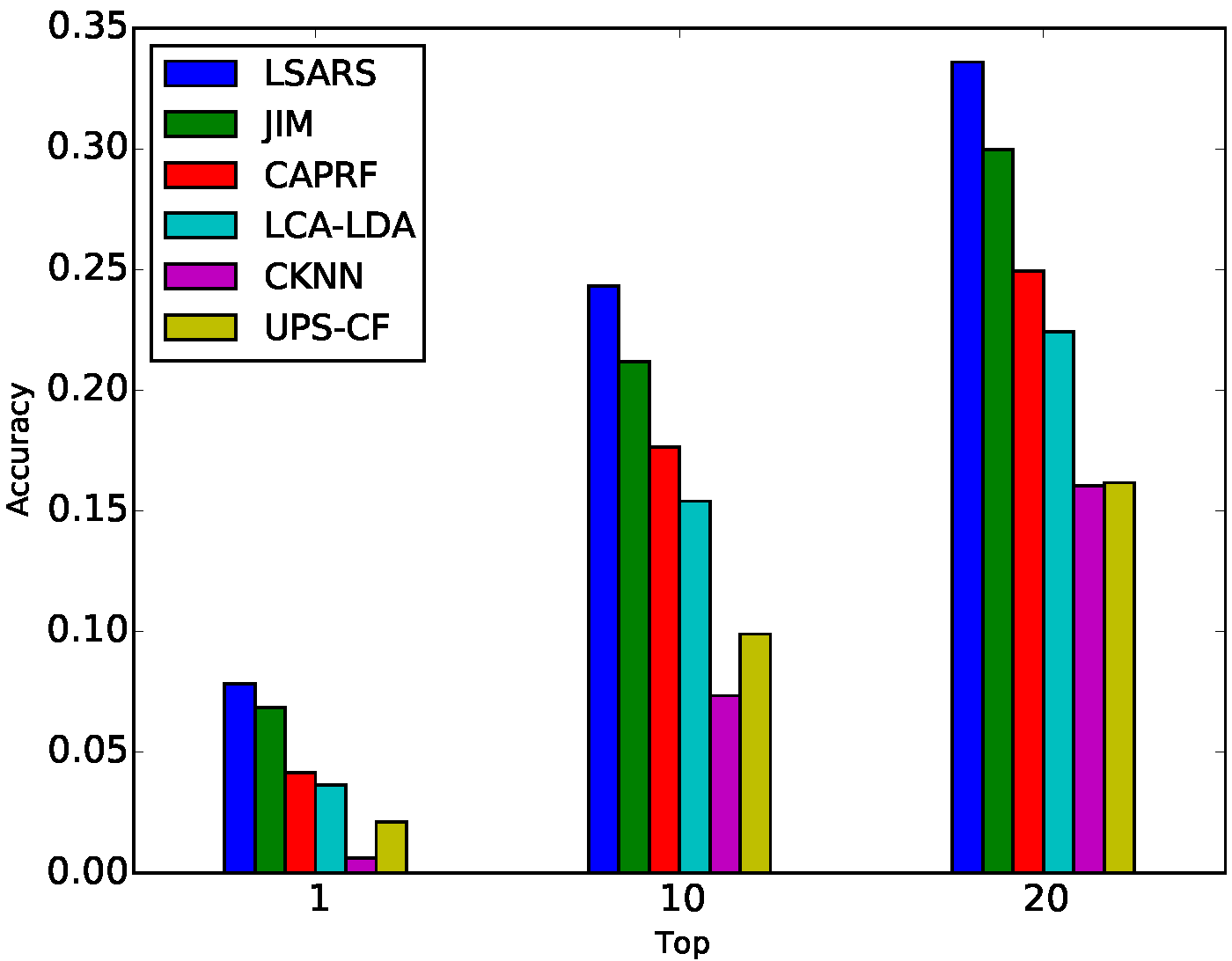}}
\caption{Spatial item recommendation on Yelp}
\label{fig:Performance on Yelp Dataset}
\end{figure}

\begin{figure}
\small
\centering
\subfigure[Home-town spatial item recommendation]{
\includegraphics[width=0.45\columnwidth]{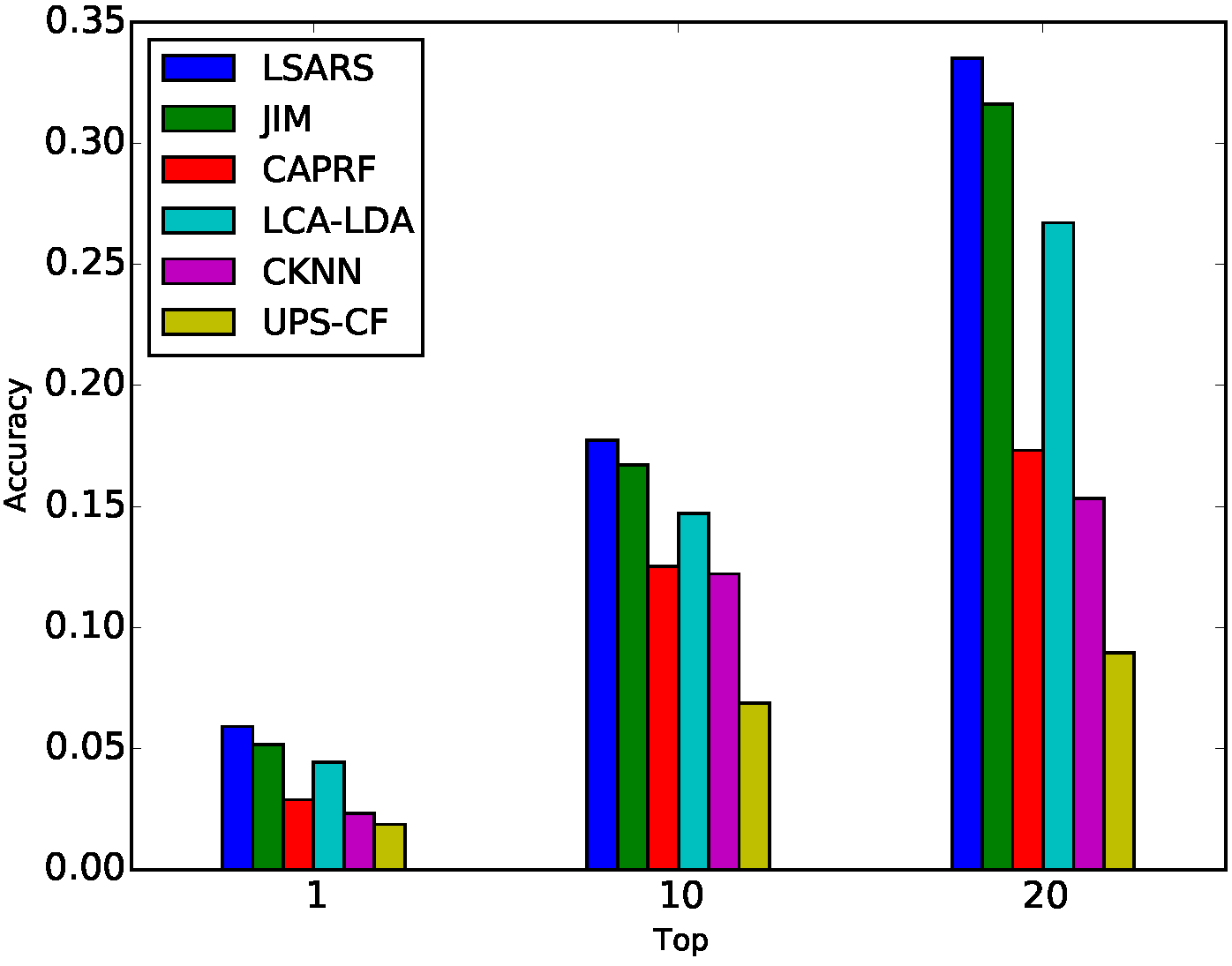}}
\hspace{0.1in}
\centering
\subfigure[Out-of-town spatial item recommendation]{
\includegraphics[width=0.45\columnwidth]{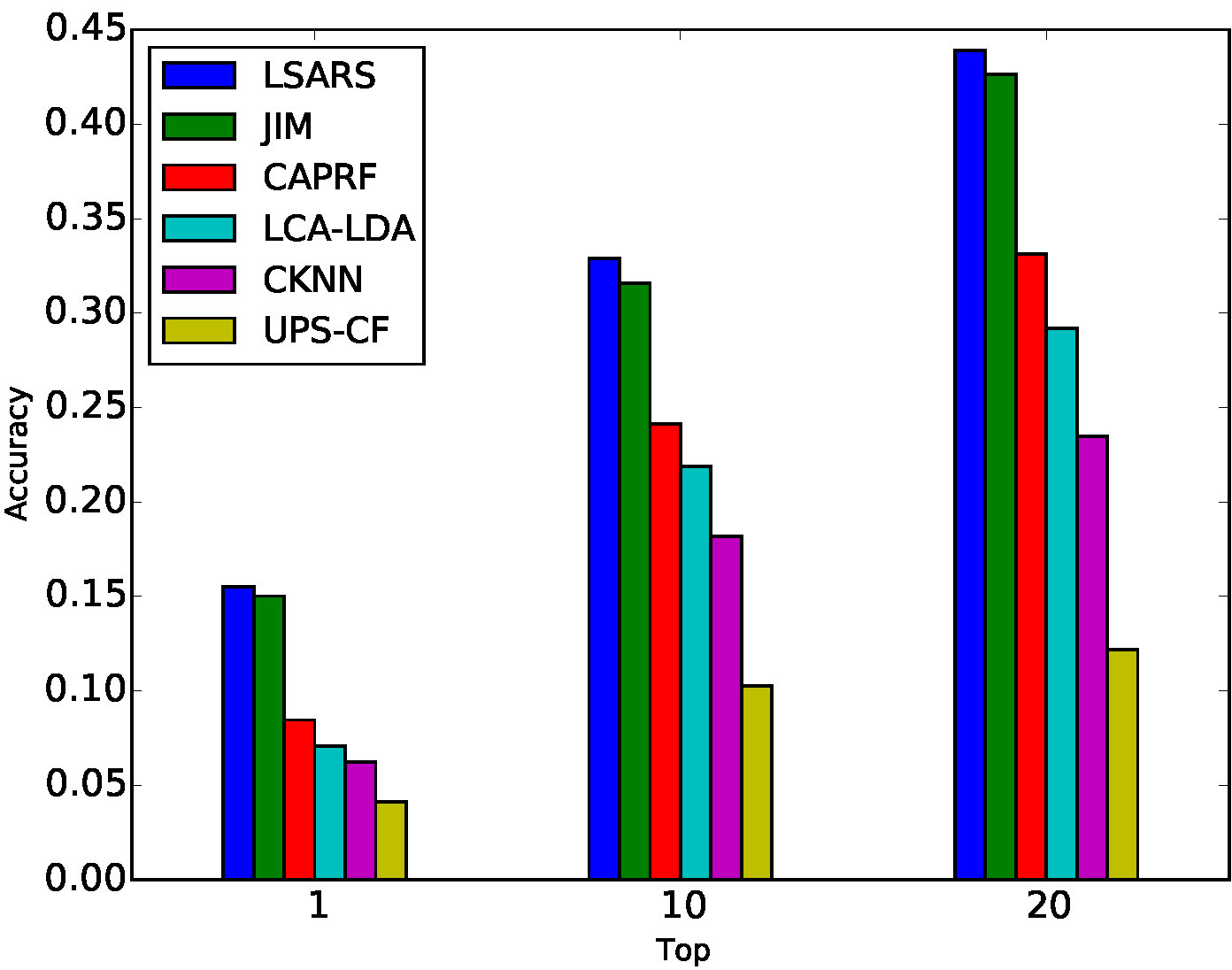}}
\caption{Spatial item recommendation on Foursquare}
\label{fig:Performance on Foursquare Dataset}
\end{figure}

Figure \ref{fig:Performance on Yelp Dataset} and \ref{fig:Performance on Foursquare Dataset} report the performance of the spatial item recommendation on the Yelp and Foursquare datasets, respectively.
We show the differences in performance when $k$ is 1, 10 and 20, as a larger value of $k$ is usually ignored for a typical top-$k$ recommendation.

It is observed that LSARS significantly outperforms the other baselines (i.e., JIM, CAPRF, LCA-LDA, CKNN and UPS-CF) on both datasets, which indicates that the recommendation accuracy can be greatly improved, by simultaneously considering the factors of geographical influence, content effect and sentiment effect. We come to the following conclusions:
1) due to the data sparsity, two CF-based methods of UPS-CF and CKNN perform worse than content-based methods of LCA-LDA, JIM and LSARS, which indicates that content information, such as reviews and descriptions of spatial items, is very valuable to alleviate the sparsity, especially for out-of-town scenario;
2) LSARS and JIM perform better than LCA-LDA, which shows that the geographical influence is useful to improve the performance;
3) LSARS outperforms CAPRF both in home-town and out-of-town recommendation scenarios, which justifies that crowd sentiment has a stronger impact on users' check-in behaviors than personal sentiment.


\begin{figure}
\small
\centering
\subfigure[Home-town target user discovery]{
\label{fig:subfig:a} 
\includegraphics[width=0.45\columnwidth]{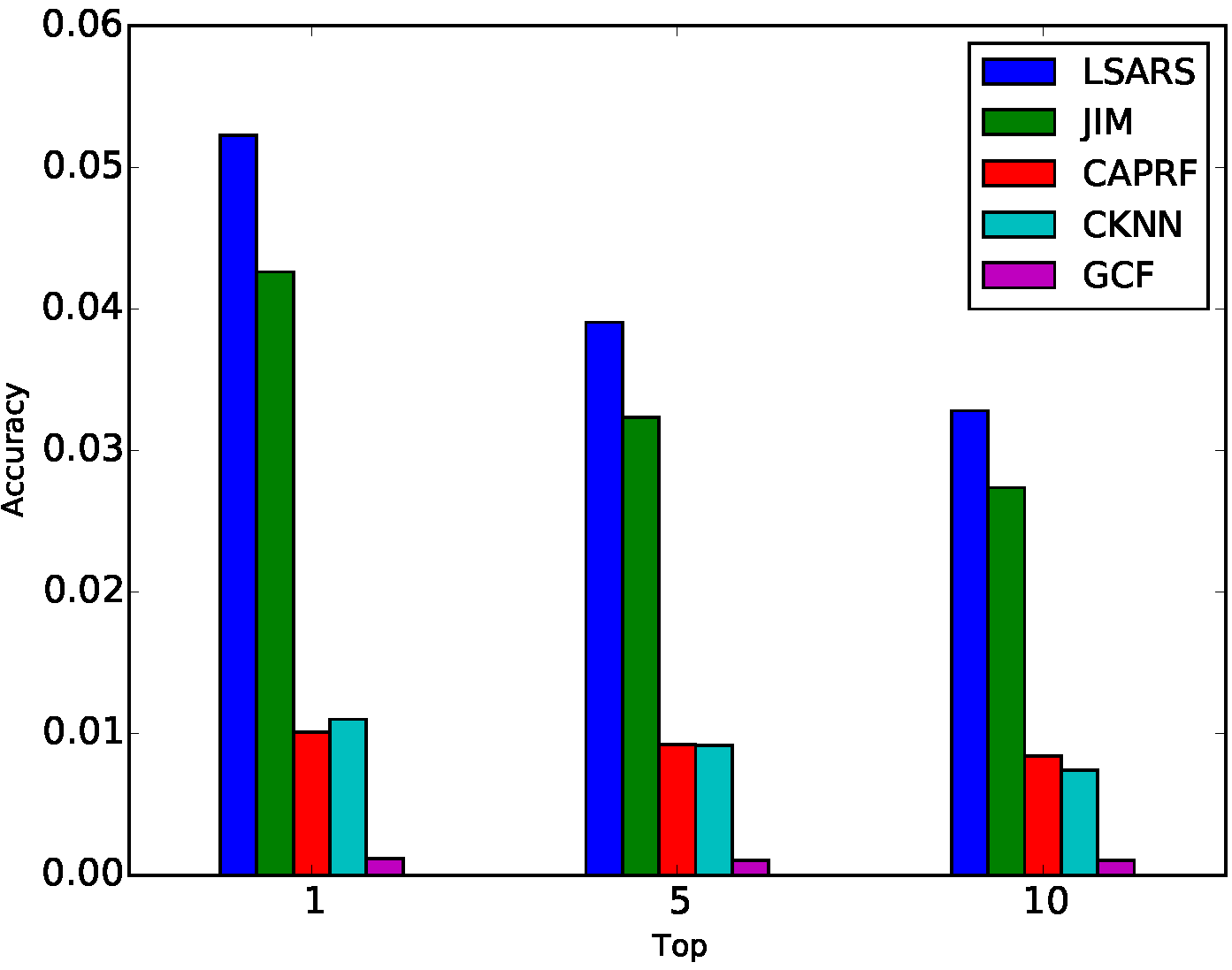}}
\hspace{0.1in}
\centering
\subfigure[Out-of-town target user discovery]{
\label{fig:subfig:b} 
\includegraphics[width=0.46\columnwidth]{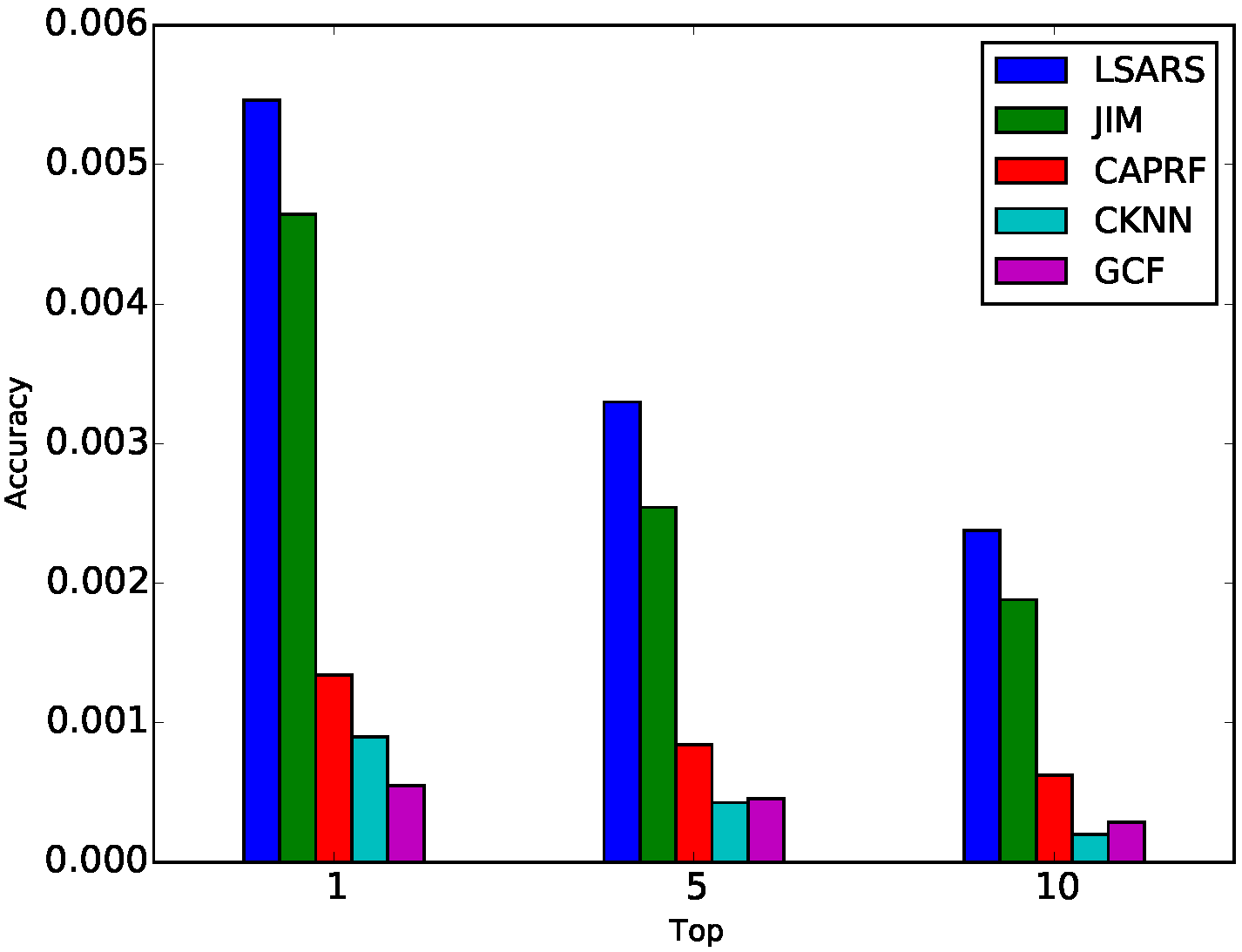}}
\caption{Target user discovery on Yelp}
\label{fig:Target User Discovery on Yelp}
\end{figure}

\begin{figure}
\small
\centering
\subfigure[Home-town target user discovery]{
\includegraphics[width=0.45\columnwidth]{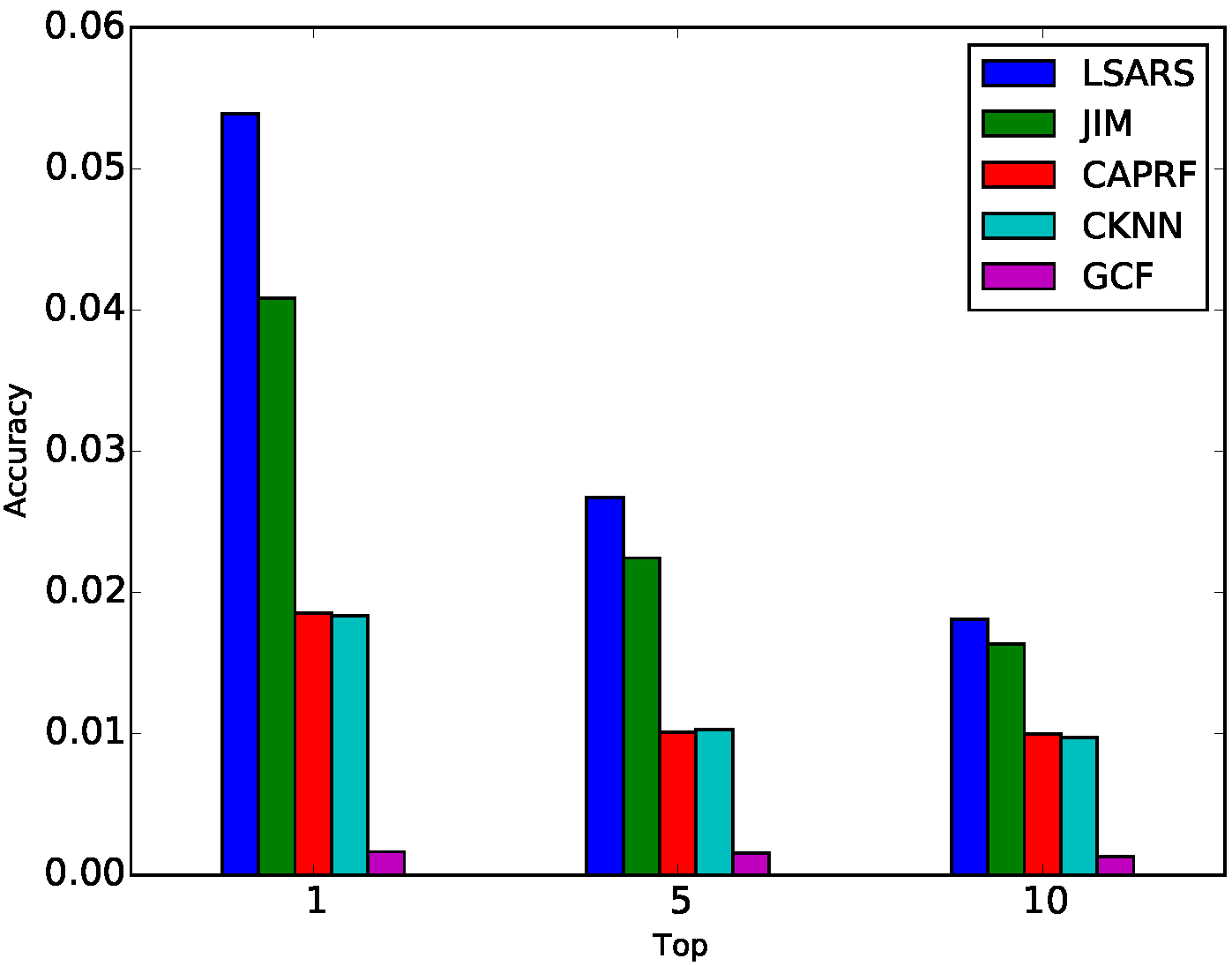}}
\hspace{0.1in}
\centering
\subfigure[Out-of-town target user discovery]{
\includegraphics[width=0.46\columnwidth]{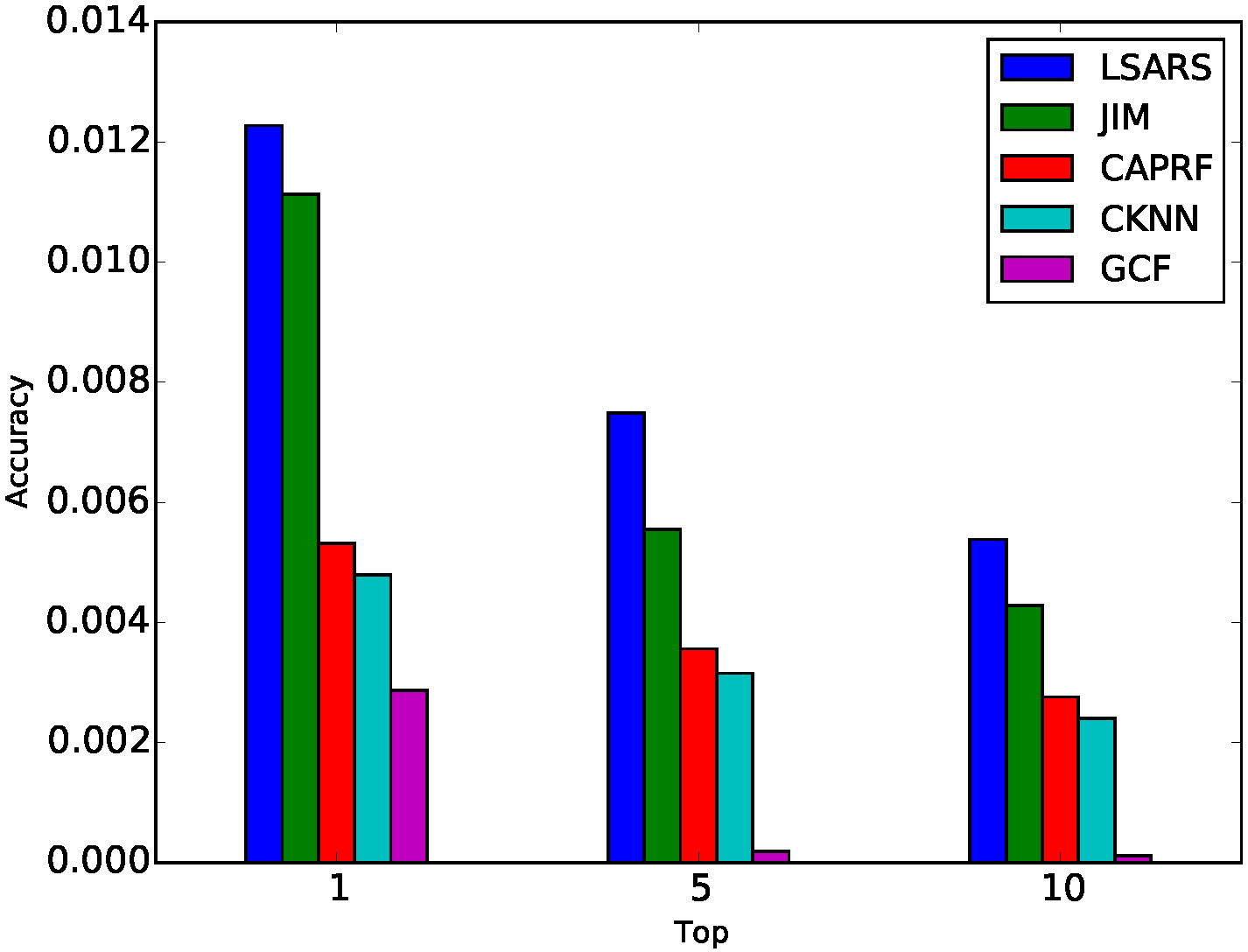}}
\caption{Target user discovery on Foursquare}
\label{fig:Target User Discovery on Foursquare}
\end{figure}

\subsubsection{Effectiveness of target user discovery }

We compare LSARS with JIM, CAPRF, GCF and CKNN for target user discovery. Note that JIM is a probabilistic model so it can be easily applied to target user discovery by simply multiplying
the conditional probability $P(v|u)$ with user popularity $P(u)$. However, CF-based methods, i.e, CAPRF, CKNN and GCF, are not originally designed for user recommendation. We simply reverse the rating matrix by treating users as items and items as users in these methods. Therefore, for CKNN, we rank all the users by the predicted rating of user $i$ on item $j$ to find the Top-$k$ users.
For CAPRF and GCF, we compute the score in the geographic preference matrix by comparing the coordinates
of the spatial items with each user' check-in records, and then normalize it over all the users.
Figure \ref{fig:Target User Discovery on Yelp} and \ref{fig:Target User Discovery on Foursquare} show the comparisons on the two datasets. Similarly, we show the performances when $k$ is set to 1, 5 and 10 since a larger \textit{k} is usually not necessary in real-world scenarios.


\vspace{-1pt}

From these figures, we can see  similar patterns of comparison results presented in the task of \emph{Spatial Item Recommendation}. The result shows that LSARS outperforms the other models (i.e., CKNN, CAPRF, GCF and JIM) on both datasets. Further observations are as below: 1) the discovery precision of out-of-town target users is much lower than that of home-town users, which indicates that the sparsity of user activity records in out-of-town regions strongly influences the performance. LSARS can perfectly solve the problem. 2) LSARS outperforms the baseline methods, which indicates that LSARS not only considers whether a spatial item satisfies users' interests, but also captures whether the item could receive a positive comment from the crowd.

\subsection{Impact of Different Parameters}

It is critial for LSARS to tune appropriate parameters for the best performance. Therefore, we study the impact of model parameters on spatial item recommendation and target user discovery on both datasets.

As for the hyperparameters $\alpha,\gamma,\beta,\delta,\eta$ and $\tau$, we let them be fixed values,
as these hyperparameters are quite independent from the other parameters.
However, the performance of LSARS is highly sensitive to the number of topics and regions.
Thus, we test the performance of LSARS by varying the number of topics and regions,
and show the results of spatial item recommendation  in Table \ref{tb:Home-town Recommendation} 
and \ref{tb:Out-of-town Recommendation Accuracy} and the results of target user discovery in
Table \ref{tb:Home-town Target User Discovery } and \ref{tb:Out-of-town Target User Discovery}.

We observe that the accuracy of spatial item recommendation first increases 
and then becomes stable as the number of topics gradually increases. However, the accuracy decreases as the number of regions increases.
Similar observations are obtained from the experimental results on the Foursquare dataset.
The optimal parameter setting for spatial item recommendation is $K = 40$ and $R = 20$.

As for target user discovery, we tune the parameters and show the results
in Table \ref{tb:Home-town Target User Discovery } and \ref{tb:Out-of-town Target User Discovery}.
The accuracy increases as the number of topics increases and then becomes stable.
On the other hand, the accuracy also increases as the number of regions increases.
Experiments on the Foursquare dataset manifest the same varying pattern.
The optimal parameter setting for target user discovery is $K = 70$ and $R = 70$.

\section{Related Work}
To improve
spatial item recommendation, some recent work has tried
to explore and integrate geo-social, temporal and semantic
information associated with users' check-in activities.

\begin{table}
\centering
\caption{Home-town recommendation Accuracy@20}
\label{tb:Home-town Recommendation}
\begin{tabular}{l|c|c|c|c|c|c}
 \hline
 \diagbox{R}{K} &K=20&K=30&K=40&K=50&K=60&K=70 \\
 \hline
 R=20& 0.326&0.325& 0.326&0.323& 0.331&0.318\\
\hline
 R=30& 0.320&0.313& 0.320&0.316& 0.319&0.317 \\
\hline
 R=40& 0.315&0.314& 0.316&0.315& 0.311&0.311 \\
\hline
 R=50& 0.307&0.306& 0.312&0.307& 0.307&0.310 \\
\hline
 R=60 & 0.302&0.301& 0.304&0.300& 0.302&0.301\\
\hline
 R=70& 0.295&0.289& 0.302&0.293& 0.301&0.292\\
 \hline
 \end{tabular}
\end{table}

\begin{table}
\centering

\caption{Out-of-town recommendation Accuracy@20}
\label{tb:Out-of-town Recommendation Accuracy}
\begin{tabular}{l|c|c|c|c|c|c}

 \hline
 \diagbox{R}{K} &K=20&K=30&K=40&K=50&K=60&K=70 \\
 \hline
 R=20& 0.320&0.333& 0.336&0.328& 0.325&0.302\\
\hline
 R=30& 0.306&0.317& 0.319&0.314& 0.302&0.314 \\
\hline
 R=40& 0.316&0.300& 0.297&0.299& 0.286&0.310 \\
\hline
 R=50& 0.290&0.279& 0.296&0.293& 0.271&0.312 \\
\hline
 R=60 & 0.266&0.277& 0.288&0.290& 0.292&0.275\\
\hline
 R=70& 0.268&0.259& 0.297&0.289& 0.283&0.296\\
 \hline
 \end{tabular}
 \end{table}


\textbf{Geo-Social Effect.} Recent studies \cite{cheng2011exploring,cho2011friendship,distance:d,ye2011exploiting,zhang2012evaluating,yin2016discovering,wang2016spore} showed that there is a strong correlation between users' check-in activities and geographical distance as well as social relationships. Thus most of these researches are mainly focused on leveraging the geographical and social influences to improve the recommendation accuracy. Ye et al. \cite{ye2011exploiting} delved into POI recommendation by investigating the geographical influences among locations and proposed a framework that combines user preferences, social influence and geographical influence. Cheng et al. \cite{cheng2012fused} investigated the geographical influence through combining a multi-center Gaussian model, matrix factorization and social influence together for location recommendation. Lian et al. \cite{lian2014geomf} incorporated spatial clustering phenomenon resulted by geographical influence into a weighted matrix factorization framework to deal with the challenge of matrix sparsity. However, all of them do not consider the role (i.e., home-town or out-of-town) of users. Thus, no matter whether users are located in the home-town regions or traveling out-of-town regions, they will be recommended the same POIs.

\textbf{Content Effect.} Most recently, researchers explored the content information of spatial items to alleviate the problem of data sparsity. Hu et al. \cite{hu2013spatial}
proposed a spatial topic model for POI recommendation considering both spatial and textual information of user posts from Twitter. To alleviate the data sparsity for out-of-town recommendation, Yin et al. \cite{topic:t2,method:m2} developed LCA-LDA and Geo-SAGE models to infer both personal interests and local preferences by exploiting the content information of visited POIs. Liu et al. \cite{liu2013point} studied the effect of content information for POI recommendation with an
integrated topic model and matrix factorization method. However, most of the above work fail to consider user interest drift when users are in the out-of-town regions.

\textbf{Sentiment Effect.} The sentiment effect of item recommendation gradually attracted much attention. Current methods combine CF and explicit sentiments to make item recommendation.\cite{garcia2013pessimists} proposed to  classify users into two distinct categories
by averaging the sentiment polarity of remarks and then use the categories as features in CF. Zhang et al. \cite{zhang2014explicit}
extracted explicit sentiments about various aspects of items from user reviews, and proposed a Explicit Factor Model (EFM) to generate explainable
recommendations. Yang et al. \cite{yang2013sentiment} proposed a hybrid user location preference model by combining location preferences, explicit sentiments and social relationships. Considering the data sparsity, some researchers used a probabilistic model to incorporate both the contents of POI and the sentiments from user reviews for POI recommendation \cite{zhao2015sar}. 

Our proposed method strategically takes user preferences, geographical influences, content effects and sentiment influences into consideration and presents a flexible probabilistic generative model for both home-town and
out-of-town recommendation. 
To deal with user interest drift, we exploit the local word-of-mouth effect according to users' current position. We model users' check-in activities in home-town and out-of-town regions by adapting to both user interest drift and crowd sentiment for spatial items.
\begin{table}
\centering
\caption{Home-town target user discovery Accuracy@5}
\label{tb:Home-town Target User Discovery }

\begin{tabular}{l|c|c|c|c|c|c}
 \hline
 \diagbox{R}{K} &K=20&K=30&K=40&K=50&K=60&K=70 \\
 \hline
 R=20& 0.026&0.026& 0.027&0.027& 0.027&0.027\\
\hline
 R=30& 0.030&0.030& 0.030&0.030& 0.030&0.030 \\
\hline
 R=40& 0.032&0.032& 0.033&0.033& 0.033&0.033 \\
\hline
 R=50&0.034&0.034& 0.035&0.035& 0.035&0.035 \\
\hline
 R=60 & 0.037&0.036& 0.036&0.036& 0.036&0.036\\
\hline
 R=70& 0.038&0.038& 0.037&0.038& 0.038&0.039\\
 \hline
 \end{tabular}

\end{table}

\begin{table}
\caption{Out-of-town target user discovery Accuracy@5}
\label{tb:Out-of-town Target User Discovery}
\begin{tabular}{l|c|c|c|c|c|c}

 \hline
 \diagbox{R}{K} &K=20&K=30&K=40&K=50&K=60&K=70 \\
 \hline
 R=20& 0.0021&0.0021& 0.0022&0.0024& 0.0025&0.0025\\
\hline
 R=30& 0.0027&0.0027& 0.0027&0.0027& 0.0028&0.0030 \\
\hline
 R=40& 0.0032&0.0032& 0.0029&0.0030& 0.0031&0.0030 \\
\hline
 R=50& 0.0033&0.0032& 0.0031&0.0031& 0.0032&0.0033 \\
\hline
 R=60 & 0.0033&0.0032& 0.0032&0.0033& 0.0032&0.0032\\
\hline
 R=70& 0.0033&0.0033& 0.0032&0.0032& 0.0033&0.0033\\
 \hline
 \end{tabular}

\end{table}

\section{Conclusions}

We propose LSARS to model users' check-in behaviors in LBSNs no matter users are in home-town or out-of-town regions, which can learn region-dependent personal interests by considering both user interest drift across geographical regions and the influence of crowd sentiment. LSARS incorporates the crowd's preferences to alleviate the data sparsity of out-of-town users' check-in records by exploiting
the local crowd's behaviors. To demonstrate its applicability and flexibility, we validate how LSARS supports two real applications in spatial item recommendation and target user discovery separately. Extensive experiments demonstrate the superiority of our proposed method over the other state-of-the-art methods.

\section{Acknowledgement}
This work is supported by National Natural Science Foundation of China (61672501, 61602453, 61402447, 61502466).
It was also partially supported by ARC Discovery Early Career Researcher Award (DE160100308), ARC Discovery Project (DP170103954).

\small
\bibliographystyle{ACM-Reference-Format}
\bibliography{sigproc}

\end{document}